\documentclass[12pt]{article}
\usepackage[left=1.0in,top=1.0in,right=1.0in,bottom=1.0in]{geometry}
\usepackage{graphicx}
\usepackage{amsmath}
\usepackage{amssymb}
\usepackage{amsthm}
\usepackage{bm}
\usepackage{color}
\usepackage{stmaryrd}
\usepackage{mathrsfs}
\usepackage{array}
\usepackage{algorithm}
\usepackage{algpseudocode,algpseudocode}
\usepackage{multirow}
\usepackage{url}
\usepackage{lineno}
\usepackage{mdframed}
\usepackage[superscript,biblabel]{cite}
\SetSymbolFont{stmry}{bold}{U}{stmry}{m}{n}

\usepackage{verbatim}

\numberwithin{equation}{section}
\newcolumntype{P}[1]{>{\centering\arraybackslash}p{#1}}
\usepackage{etoolbox}

\AfterEndEnvironment{assumption}{\noindent\ignorespaces}

\makeatletter
\newcommand*{\rom}[1]{\expandafter\@slowromancap\romannumeral #1@}
\makeatother

\begin{document}

\begin{center}

    \vspace*{1cm}
   
    \textbf{Validation of the Reduced Unified Continuum Formulation Against \\ In Vitro 4D-Flow MRI}
    
    \vspace{1cm}
    
    \textbf{Ingrid S. Lan$^{1}$, Ju Liu$^{2,3}$, Weiguang Yang$^{4}$, Judith Zimmermann$^{5,6}$, \\ Daniel B. Ennis$^{5,7}$, Alison L. Marsden$^{1,4,8}$}
    
    \vspace{1.5cm}
    \textbf{Corresponding Author:} Alison L. Marsden \\
    \textbf{(e)} amarsden@stanford.edu \\
     
    \vspace{1cm}
    \begin{enumerate}
    \item Department of Bioengineering, Stanford University, Stanford, CA 94305, USA
    \item Department of Mechanics and Aerospace Engineering, Southern University of Science and Technology, Shenzhen, Guangdong 518055, P.R. China
    \item Guangdong-Hong Kong-Macao Joint Laboratory for Data-Driven Fluid Mechanics and Engineering Applications, Southern University of Science and Technology, Shenzhen, Guangdong 518055, P.R. China
    \item Department of Pediatrics (Cardiology), Stanford University, Stanford, CA 94305, USA
    \item Department of Radiology, Stanford University, Stanford, CA 94305, USA
    \item Department of Informatics, Technical University of Munich, 85748 Garching, Germany
    \item Division of Radiology, Veterans Affairs Health Care System, Palo Alto, CA 94304, USA
    \item Institute for Computational and Mathematical Engineering, Stanford University, Stanford, CA 94305, USA
    \end{enumerate}

\end{center}

\newpage

\section*{Abstract}
In our recent work, we introduced the reduced unified continuum formulation for vascular fluid-structure interaction (FSI) and demonstrated enhanced solver accuracy, scalability, and performance compared to conventional approaches. We further verified the formulation against Womersley's deformable wall theory. In this study, we assessed its performance in a compliant patient-specific aortic model by leveraging 3D printing, 2D magnetic resonance imaging (MRI), and 4D-flow MRI to extract high-resolution anatomical and hemodynamic information from an in vitro flow circuit. To accurately reflect experimental conditions, we additionally enabled in-plane vascular motion at each inlet and outlet, and implemented viscoelastic external tissue support and vascular tissue prestressing. Validation of our formulation is achieved through close quantitative agreement in pressures, lumen area changes, pulse wave velocity, and early systolic velocities, as well as qualitative agreement in late systolic flow structures. Our validated suite of FSI techniques can be used to investigate vascular disease initiation, progression, and treatment at a computational cost on the same order as that of rigid-walled simulations. This study is the first to validate a cardiovascular FSI formulation against an in vitro flow circuit involving a compliant vascular phantom of complex patient-specific anatomy. \\

\noindent \textbf{Key Terms:}  Fluid-structure interaction, Pulse wave velocity, Magnetic resonance imaging, Compliant 3D printing, In vitro validation

\vspace{1.5cm}

\noindent \textbf{Abbreviations}
\begin{table*}[htbp]
\begin{center}
\tabcolsep=0.25cm
\renewcommand{\arraystretch}{1.2}
\begin{tabular}{c | c}
CFD & computational fluid dynamics \\
CMM & coupled momentum method \\
FSI & fluid-structure interaction \\
GRE & gradient echo \\
LSE & least squares error \\
PC-MRI & phase contrast magnetic resonance imaging \\
PWV & pulse wave velocity \\
RANSAC & Random Sample Consensus \\
RUC & reduced unified continuum \\
SPGR & spoiled gradient echo MRI \\
TTF & time-to-foot
\end{tabular}
\end{center}
\end{table*}

\newpage

\noindent \textbf{Glossary of Terms}
\begin{table*}[htbp]
\begin{center}
\tabcolsep=0.25cm
\renewcommand{\arraystretch}{1.05}
\begin{tabular}{c c c c c}
Symbol & Name & Definition & SI Unit \\
\hline
$\Omega$ & FSI domain & FSI domain & - \\
$\Omega^f$ & fluid domain & fluid domain & - \\
$\Omega^s$ & solid domain & solid domain & - \\
$\Gamma_I$ & fluid-solid interface & interface between $\Omega^s$ and $\Omega^f$ & - \\
$\Gamma_W$ & solid outer wall & outer wall of $\Omega^s$ & - \\
$\Gamma^{i}_{\mathrm{cap}}$ & $i$-th annular solid cap & $i$-th annular solid cap & - \\
$\Gamma^{i}_{\mathrm{ring}}$ & $i$-th solid ring & $i$-th solid ring on $\Gamma_I$& - \\
$\bm n^s$ & solid unit outward normal & unit outward normal vector of $\Omega^s$ & - \\
$\bm u^s$ & solid displacement & solid displacement & m \\
$\bm u^w$ & membrane displacement & membrane displacement on $\Gamma_I$ & m \\
$\bm v^s$ & solid velocity & solid velocity & m/s \\
$\rho^s$ & solid density & solid density & kg/m$^3$ \\
$\bm \sigma^s$ & solid Cauchy stress & solid Cauchy stress & N/m$^2$ \\
$\bm \sigma_0$ & solid prestress & $\bm \sigma^s$ at imaging & N/m$^2$ \\
$\bm b^s$ & solid body force & solid body force & N/kg \\
$\bm \sigma^{s, l}$ & solid lamina Cauchy stress & $\bm \sigma^s$ in lamina coordinate system & N/m$^2$ \\
$\bm u^{s,l}$ & solid lamina displacement & $\bm u^s$ in lamina coordinate system & m \\
$E$ & Young's modulus & Young's modulus & N/m$^2$ \\
$\nu$ & Poisson's ratio & Poisson's ratio & - \\
$k^s$ & spring constant & external elastic support & kg/(m$^2 \cdot$ s$^2$) \\
$c^s$ & damping constant & external viscous support & kg/(m$^2 \cdot$ s) \\
$h^s$ & wall thickness & wall thickness & m \\
$\bm n^f$ & fluid unit outward normal & unit outward normal vector of $\Omega^f$ & - \\
$\bm v^f$ & fluid velocity & fluid velocity & m/s \\
$p^f$ & fluid pressure & fluid pressure & N/m$^2$ \\
$\rho^f$ & fluid density & fluid density & kg/m$^3$ \\
$\mu^f$ & fluid dynamic viscosity & fluid dynamic viscosity & N $\cdot$ s/m$^2$\\
$\bm b^f$ & fluid body force & fluid body force & N/kg \\
$\bm h^f$ & fluid boundary traction & fluid boundary traction & N \\
$T_p$ & cardiac period & length of cardiac cycle & s
\end{tabular}
\end{center}
\end{table*}

\newpage

\section{Introduction}
\label{sec:introduction}

As image-based computational fluid dynamics (CFD) and fluid-structure interaction (FSI) simulations continue to gain traction for predictive and personalized medicine, it is imperative that numerical CFD and FSI methods be verified and, furthermore, validated against in vivo and/or in vitro data. Phase contrast magnetic resonance imaging (PC-MRI), which encodes absolute velocities of coherent blood flow, was previously limited to unidirectional velocity encoding in 2D. Recent developments in 4D-flow MRI, however, have enabled three-directional velocity encoding over 3D volumes, paving the way for increasingly detailed numerical validation.

We recently developed a unified continuum formulation for FSI \cite{Liu2018} that not only recovers important continuum models including viscous fluids and visco-hyperelatic solids \cite{Liu2021}, but is also well-behaved in both compressible and fully incompressible regimes. This unified continuum formulation was then simplified to a reduced unified continuum (RUC) formulation via consideration of three modeling assumptions for vascular FSI, namely the infinitesimal strain, thin-walled, and membrane assumptions \cite{Lan2021}. The RUC formulation, which achieves monolithic coupling of the fluid and solid subproblems in an Eulerian frame, was found to offer computational cost as low as $1.3$ times that of rigid-walled CFD simulations \cite{Lan2022}. While the coupled momentum method (CMM) \cite{Figueroa2006} similarly embeds a linear elastic membrane into an Eulerian fluid subproblem, key theoretical and numerical differences exist with regard to the fluid-solid coupling, spatiotemporal discretization, vascular wall dynamics, and linear solver technology \cite{Lan2021}. In our verification against Womersley's deformable wall theory, we demonstrated notable agreement between analytical and numerical solutions. Given the overlap in assumptions in the RUC formulation and Womersley's deformable wall theory, assessing the validity of our adopted assumptions in settings of practical clinical interest, particularly with complex anatomical geometries, remains necessary.

CFD simulations of the cardiovascular system have commonly been validated against in vivo velocities from either 2D cine PC-MRI \cite{Long2000, Ku2002, Cheng2014} or 4D-flow MRI \cite{Saitta2019, Pons2020, Annio2021}. Numerical validation can alternatively be performed against in vitro MRI of experimental flow phantoms embedded in benchtop circuits to assess effects of various parameters, including boundary conditions, anatomical geometries, and wall mechanical properties, on the flow behavior and solver performance. Constrained by limitations in fabrication methods, these in vitro experiments have until recently employed rigid flow phantoms constructed from photoreactive resin \cite{Kung2011a, Zhou2015, Biglino2015, Kaiser2021}. Alastruey et al. \cite{Alastruey2006} and Kung et al. \cite{Kung2011} were among the first to perform in vitro FSI validation on compliant flow phantoms of idealized geometries that were fabricated from silicone dip-spin coating or hand-painting. Numerous other flow circuits with compliant flow phantoms fabricated from silicone, polyurethane, or latex \cite{Tanne2010, Kolyva2012, Knoops2017} have also been engineered to investigate cardiovascular hemodynamics in health and disease and to further assess the performance of implantable devices. Recent advances in 3D printing techniques \cite{Ionita2014}, including PolyJet and stereolithography, now enable rapid, repeatable printing of compliant patient-specific flow phantoms from novel photopolymers \cite{Biglino2013, Ho2020, Zimmermann2021} without the need for laborious procedures. Importantly, mechanical characterization of these cost-effective phantoms can be performed to inform FSI validation studies. Nonetheless, to our knowledge, no previous studies have validated cardiovascular FSI formulations against in vitro flow circuits involving compliant patient-specific vascular phantoms.

In a previously published study \cite{Zimmermann2021}, we demonstrated the use of novel compliant 3D printing to fabricate patient-specific aortic phantoms of three stiffness values, which were then embedded in an MRI-compatible flow circuit under physiological hemodynamic conditions. In our current study, we focus only on the most compliant phantom and assess the RUC formulation by drawing direct comparisons to the experimentally measured three-component 3D velocities, flow rates, pressures, luminal area changes, and pulse wave velocity. Effects of selected boundary conditions are also assessed.

\section{Materials and Methods}
In this section, we summarize the experimental methods \cite{Zimmermann2021} adopted to acquire the in vitro MRI data as well as the numerical methods employed to simulate the flow circuit.

\subsection{3D-Printed Aortic Flow Phantom}
Under a protocol approved by the Stanford Institutional Review Board, an in vivo chest 4D-flow MRI of a 50-year-old male subject was acquired with informed consent and subsequently used to construct a 3D anatomical model of the thoracic aorta (\textbf{Figure \ref{fig:experimental-setup}A}) in the open-source software package SimVascular \cite{Lan2018}. We modeled the ascending aortic inlet (\textit{inlet}) and four outlets, namely the brachiocephalic artery (\textit{BCA}), left common carotid artery (\textit{LCA}), left subclavian artery (\textit{LSA}), and descending aorta (\textit{outlet}). To define the outer wall surface, Meshmixer (Autodesk) was used to extrude surface mesh nodes by the wall thickness $h^s = 0.2$ cm. Each inlet and outlet was finally extended by $2$ cm to facilitate tubing connections in the flow circuit.

A PolyJet photopolymerization 3D printer (Stratasys) was used to print the phantom from a material blend of the Agilus30 and VeroClear photopolymers (\textbf{Figure \ref{fig:experimental-setup}B}). Three dumbbell-shaped samples were additionally printed for uniaxial tensile testing to 50\% peak strain at a strain rate of 25\% s$^{-1}$. The tangential Young's modulus at a nominal stress corresponding to the experimentally measured mean pressure ($56$ mm Hg) was determined to be $1.27 \times 10^7$ dyn/cm$^2$.

\subsection{MRI-Compatible In Vitro Flow Circuit}
\label{subsec:flow-circuit}
\begin{figure}
\begin{center}
\begin{tabular}{c}
\includegraphics[angle=0, trim=75 115 70 100, clip=true, scale=0.9]{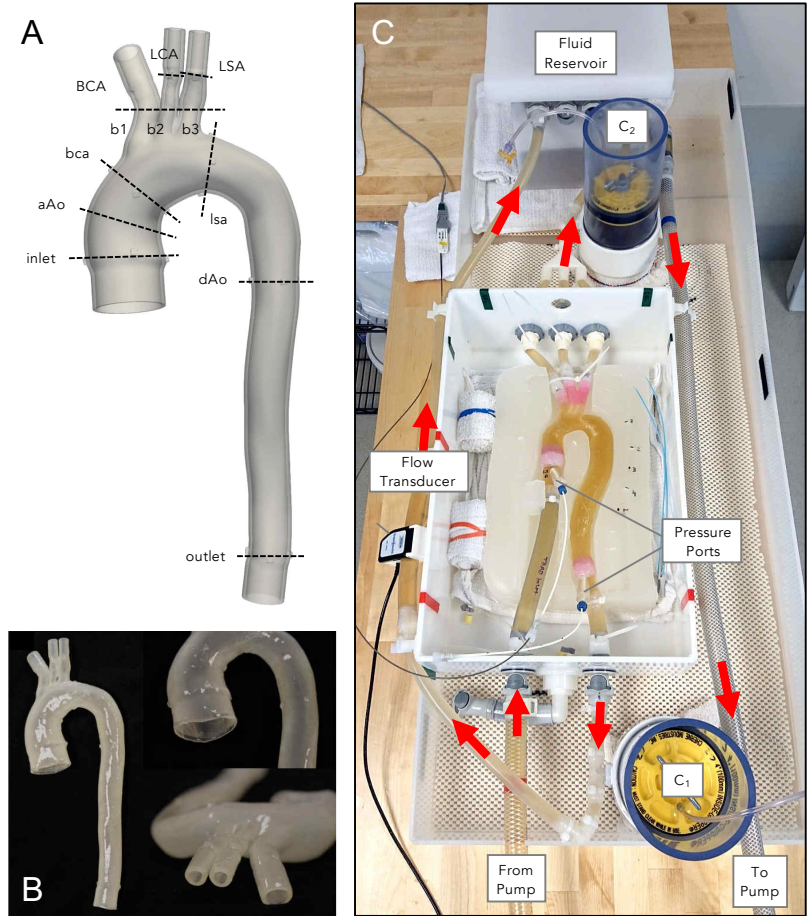}
\end{tabular}
\end{center}
\caption{(A) Print-ready 3D STL model of a patient-specific thoracic aorta, annotated with the caps and landmark slices for 2D cine PC-MRI and 2D cine GRE MRI. (B) The resulting compliant 3D-printed flow phantom with $2$-cm extensions on all five caps to facilitate connection to custom barbed model-tubing transition elements. (C) MRI-compatible in vitro flow circuit consisting of a programmable flow pump, a fluid reservoir, tubing with pinch valves serving as resistance elements, and two sealed air compression chambers ($C_1$, $C_2$) serving as capacitance elements. A flow transducer and two pressure transducers were inserted for resistance and capacitance tuning prior to transfer of the flow circuit into the MRI scanner. Red arrows indicate the direction of flow.} 
\label{fig:experimental-setup}
\end{figure}

The 3D-printed aortic flow phantom was embedded in a gel block (\textbf{Figure \ref{fig:experimental-setup}C}) to ensure repeatable positioning and to provide a static ``tissue" reference for eddy current phase offset correction. Components of the in vitro flow circuit included an MRI-compatible programmable flow pump (Shelley Medical Imaging Technologies), a fluid reservoir, sealed air compression chambers serving as capacitance elements, and pinch valves distal to the outlets to model distal vascular resistance. A 40\%-60\% glycerol-water mixture was used to mimic blood density and viscosity. Pressure transducers (Millar) were inserted at \textit{inlet} and \textit{outlet}, and an ultrasonic flow probe (Transonic Systems) was clamped at \textit{outlet}. Resistance and capacitance parameters were tuned to achieve physiological pressures and flow splits prior to removal of the pressure and flow transducers.

\subsection{MRI Acquisitions}
All imaging experiments were performed with a 3 Tesla MRI scanner (Siemens Healthineers) at a temporal resolution of $0.02$ s. 3D spoiled gradient echo (SPGR) was first performed at a steady flow of $71.2$ mL/s to capture high-resolution anatomical information. Subsequently, 2D cine PC-MRI, 2D cine gradient echo (GRE) MRI, and 4D-flow MRI were performed under pulsatile flow, programmed to achieve mean and peak flow rates of $71.2$ mL/s and $300$ mL/s, respectively, at a cardiac period $T_p = 1.0$ s. The $7$ planes at which 2D cine PC-MRI and 2D cine GRE MRI were performed (\textbf{Figure \ref{fig:experimental-setup}A}) corresponded to the ascending aortic inlet (\textit{inlet}), ascending aorta (\textit{aAo}), arch proximal to the brachiocephalic artery (\textit{bca}), the three neck arteries (\textit{b1}, \textit{b2}, \textit{b3}), arch distal to the left subclavian artery (\textit{lsa}), mid-descending aorta (\textit{dAo}), and the descending aortic outlet (\textit{outlet}). 4D-flow scans were corrected for artifacts produced by Maxwell terms, gradient nonlinearity, and eddy currents.

\subsection{Image Analysis}
\label{subsec:experimental-analysis}
Time-varying lumen contours were automatically tracked in the 2D cine GRE scans. Time-varying 2D cine PC-MRI velocities were then masked with these lumen contours and integrated into flow rate waveforms.

To determine the pulse wave velocity (PWV), we analyzed temporal shifts in the time-to-foot (TTF) \cite{Markl2010} of flow rate waveforms. Specifically, lumen contours at $50$ equidistant normal slices along the descending aortic centerline (from \textit{lsa} to \textit{outlet}) were automatically tracked in the 4D-flow magnitude images. Time-varying 4D-flow MRI velocities were then masked with these lumen contours and integrated into flow rate waveforms. The TTF for each flow rate waveform was determined to be the time of intersection between the following two lines: (i) horizontal line through the waveform's diastolic value, and (ii) line through the waveform's upslope points at 20\% and 80\% of the way from the diastolic value to its peak flow rate. The PWV was finally defined to be the slope of the linear regression line fitted to all $50$ TTFs plotted against the corresponding centerline positions. In addition to a least squares error (LSE) regression, a linear regression was also performed using Random Sample Consensus (RANSAC) to exclude outliers.

\subsection{Computational Model and Mesh Generation}
We employed SimVascular to segment the steady deformed configuration of the aortic phantom from the 3D SPGR scan. The inlet and outlets were truncated to exclude the artificial $2$-cm cap extensions (\textbf{Figure \ref{fig:simulation-methods}}). TetGen was used to discretize the 3D anatomical model with linear tetrahedral elements and three boundary layers at a layer decreasing ratio of $0.5$. Using meshes up to $3.92 \times 10^6$ elements, a mesh convergence study was performed for steady-state diastolic simulations. Based on a tolerance criterion of $3\%$ variation across all cap pressures and flows, we settled on a mesh with $1.98 \times 10^6$ elements and $3.35 \times 10^5$ nodes.

\subsection{RUC Formulation for Vascular FSI}

\subsubsection{Strong-Form FSI Problem}
We consider a domain $\Omega \subset \mathbb R^3$ admitting a non-overlapping subdivision $\overline{\Omega} = \overline{\Omega^f \cup \Omega^s}$, $\emptyset = \Omega^f \cap \Omega^s$, in which $\Omega^f$ and $\Omega^s$ represent the fluid and solid subdomains with unit outward normal vectors $\bm n^f$ and $\bm n^s$, respectively. The fluid-solid interface $\Gamma_I$ is a two-dimensional manifold on which $\bm n^f = - \bm n^s$.

Under the infinitesimal strain assumption, the solid governing equations posed in $\Omega^s$ are as follows,
\begin{align*}
& \bm 0 = \frac{d \bm u^s}{d t} - \bm v^s, \\
& \bm 0 = \rho^s \frac{d \bm v^s}{d t} - \nabla \cdot \bm \sigma^s - \rho^s \bm b^s,
\end{align*}
where $\bm u^s$, $\bm v^s$,  $\rho^s$, $\bm \sigma^s$, and $\bm b^s$ are the solid displacement, velocity, density,  Cauchy stress, and body force per unit mass, respectively. Considering isotropic linear elasticity, we express the constitutive relation as follows in the lamina coordinate system,
\begin{align*}
& \bm \sigma^{s, l} = \mathbb C^{s, l}  \bm \epsilon^{l}(\bm u^{s,l}), \quad \mathbb C^{s, l} := 2 \mu^s(\bm x^l) \mathbb I + \lambda^s(\bm x^l) \bm I \otimes \bm I, \quad \bm \epsilon^{l}(\bm u^{s,l}) := \frac12 \left( \nabla \bm u^{s,l} + \left( \nabla \bm u^{s,l} \right) ^T \right), 
\end{align*}
where $\bm \sigma^{s, l}$ and $\bm \epsilon^{l}(\bm u^{s,l})$ are respectively the Cauchy stress and infinitesimal strain in the lamina coordinate system, $\bm I$ is the second-order identity tensor, $\mathbb I$ is the fourth-order symmetric identity tensor, and $\mu^s$ and $\lambda^s$ are the Lam\'e parameters. In Voigt notation,
\begin{align*}
& \bm \sigma^{s, l} = \left[ \sigma_{11}^{s,l}, \sigma_{22}^{s,l}, \sigma_{12}^{s,l}, \sigma_{23}^{s,l} , \sigma_{31}^{s,l} \right]^T, \displaybreak[2] \\ 
& \bm \epsilon^{l}(\bm u^{s, l}) = \left[ \epsilon_{11}^{l}, 
\epsilon_{22}^{l}, 2 \epsilon_{12}^{l}, 2 \epsilon_{23}^{l}, 2 \varepsilon_{31}^{l} \right]^T = \left[ u_{1,1}^{s,l} , u_{2,2}^{s,l}, u_{1,2}^{s,l} + u_{2,1}^{s,l}, u_{3,2}^{s,l}, u_{3,1}^{s,l} \right]^T, \displaybreak[2] \\
& \mathbb C^{s, l} = \frac{E}{(1 - \nu^2)}
\begin{bmatrix}
1 & \nu &  & &  \\[1mm]
\nu & 1 & &  &  \\[1mm]
 &  & \displaystyle \frac{1 - \nu}{2} & &  \\[1mm]
 &  &  & \kappa \displaystyle \frac{(1 - \nu)}{2} &  \\[1mm]
 &  &  &  & \kappa \displaystyle \frac{ (1 - \nu)}{2} \\[1mm]
\end{bmatrix},
\end{align*}
where $E$ is the Young's modulus, $\nu$ is the Poisson's ratio, and $\kappa = 5/6$ is the shear correction factor. Experimentally determined wall properties were uniformly prescribed: $E = 1.27 \times 10^7$ dyn/cm$^2$, $h^s = 0.2$ cm, $\rho^s = 1.0$ g/cm$^3$, $\nu = 0.5$. Using a rotation matrix $\bm Q$ transforming global coordinates to lamina coordinates, we then compute the Cauchy stress in the global coordinate system as $\bm \sigma^s = \bm Q^T \bm \sigma^{s,l} \bm Q$.

In the fluid subdomain $\Omega^f$, we consider an incompressible Newtonian fluid governed by the following equations in an Eulerian frame,
\begin{align*}
& \bm 0 = \rho^f \frac{\partial \bm v^f}{\partial t}  + \rho^f  \bm v^f \cdot \nabla \bm v^f - \nabla \cdot \bm \sigma^f_{\mathrm{dev}} + \nabla p^f - \rho^f \bm b^f, \\
& 0 = \nabla \cdot \bm v^f,
\end{align*}
where $\bm v^f$, $p^f$, $\rho^f$, $\mu^f$, and $\bm b^f$ are the fluid velocity, pressure, density, dynamic viscosity, and body force per unit mass, respectively; $\bm \sigma^f_{\mathrm{dev}}$ and $\bm \varepsilon_{\mathrm{dev}}$ are the deviatoric parts of the Cauchy stress and rate-of-strain,
\begin{align*}
& \bm \sigma^f_{\mathrm{dev}} := 2\mu^f \bm \varepsilon_{\mathrm{dev}}(\bm v^f), \quad \bm \varepsilon_{\mathrm{dev}}(\bm v^f) := \frac12 \left( \nabla \bm v^f + \left( \nabla \bm v^{f} \right)^T \right) - \frac13 \nabla \cdot \bm v^f \bm I,
\end{align*}
and the fluid Cauchy stress is given by $\bm \sigma^f := \bm \sigma^f_{\mathrm{dev}} -p^f \bm I$. In this work, we set the fluid density $\rho^f$ to $1.06$ g/cm$^3$ and fluid viscosity $\mu^f$ to $0.04$ Poise.

The strong-form FSI problem is completed with the kinematic and dynamic coupling conditions enforcing the continuity of velocity and traction on $\Gamma_I$, respectively,
\begin{align*}
\bm v^f = \bm v^s, \qquad \bm \sigma^f \bm n^f = -\bm \sigma^s \bm n^s.
\end{align*}

\subsubsection{Numerical Boundary Conditions}
\label{subsec:bcs}
The boundary of the solid subdomain can be decomposed as $\partial \Omega^s = \Gamma_{I} \cup \Gamma_{W} \cup \bigcup_{i=1}^{\mathrm n_{\mathrm{cap}}}\Gamma^{i}_{\mathrm{cap}}$, where $\Gamma_{W}$ represents the outer wall, $\Gamma^{i}_{\mathrm{cap}}$ represents the $i$-th annular cap surface, and $\mathrm n_{\mathrm{cap}}=5$ represents the number of cap surfaces. While $\Gamma_{W}$ has conventionally been modeled as a stress-free surface, a Robin boundary condition was recently introduced to represent the viscoelastic behavior of tissues and organs surrounding the modeled vasculature \cite{Moireau2012}. To model the surrounding gel block, we imposed $\bm \sigma^s \bm n^s = -k^s \bm u^s - c^s \bm v^f$ on $\Gamma_{W}$, where the spring constant $k^s = 0$ g/(cm$^2 \cdot$ s$^2$) and damping constant $c^s = 3.0 \times 10^5$ g/(cm$^2 \cdot$ s) were determined to best match experimental relative luminal area changes.

Furthermore, most vascular FSI simulations have enforced \textit{clamping}, or zero displacements, at all inlets and outlets of the solid subdomain, yet the phantom inlet and outlets here were not experimentally clamped, given the $2$-cm extensions beyond the domain of interest. More recently, studies have relaxed this displacement constraint to enable either purely radial motion \cite{Fonken2021} or in-plane motion \cite{Bazilevs2010}. We similarly enabled in-plane motion on all caps by applying inclined `roller' boundary conditions on the associated displacements as follows,
\begin{align*}
& \bm u^s \cdot \bm n^s = 0 \quad \mbox{ and } \quad \bm \sigma^s \bm n^s - \left( \bm \sigma^s \bm n^s \cdot \bm n^s \right) \bm n^s = \bm 0, \quad \mbox{ on } \Gamma^{i}_{\mathrm{cap}} \quad \mbox{ for } i = 1, \cdots, 5.
\end{align*}
To constrain degrees of freedom in \textit{skew directions} \cite{Griffiths1990}, a \textit{skew} coordinate system was established for each cap using $\bm n^s$ and two orthonormal vectors tangential to $\Gamma^{i}_{\mathrm{cap}}$. All associated equations were solved in the \textit{skew} coordinate system, and solutions for the associated degrees of freedom were subsequently rotated back into the global coordinate system.

The boundary of the fluid subdomain can be decomposed as $\partial \Omega^f = \Gamma_{I} \cup \bigcup_{k=1}^{\mathrm n_{g}} \Gamma^{k}_{g} \cup \bigcup_{l=1}^{\mathrm n_{h}} \Gamma^{l}_{h}$, where boundary surface $\Gamma^{k}_{g}$ represents the $k$-th surface prescribed with Dirichlet data $\bm g^k(\bm x, t)$ as $\bm v^f = \bm g^k(\bm x, t)$, and $\Gamma^{l}_{h}$ represents the $l$-th surface prescribed with Neumann data $-P^{l}(t)$ as $\bm \sigma^f \bm n^f = -P^{l}(t)\bm n^f$. Here, we imposed Dirichlet conditions on \textit{inlet} and the three neck artery outlets (\textit{BCA}, \textit{LCA}, and \textit{LSA}) and a Neumann condition on \textit{outlet}, yielding $\mathrm n_{g}=4$ and $\mathrm n_{h}=1$. In the remainder of this work, we eliminate the superscript $l$ for the Neumann boundary to simplify notation.

To prescribe time-varying velocity fields on \textit{inlet}, we leveraged the three-component velocities from 4D-flow MRI rather than the unidirectional, through-plane velocities from 2D cine PC-MRI. Given the different coordinate systems across the various imaging sequences, 4D-flow and 2D cine GRE scans were first registered to the computational mesh via rigid coherent point drift. Time-varying \textit{inlet} lumen contours from 2D cine GRE were then used to mask the 4D-flow velocities. To map these deforming luminal domains onto the stationary inlet surface mesh, a piecewise linear transformation was defined to map each experimental lumen contour to the boundary of the inlet mesh. The same piecewise linear transformation was then applied to the luminal domain of the masked inlet velocity profile, and the resulting velocity profile was interpolated onto the inlet mesh with a Gaussian kernel. Linear interpolation of the velocity fields was performed between consecutive temporal frames.

Despite use of the gel block to ensure repeatable phantom positioning, deformations were experimentally introduced, given the small size of the neck arteries and compliance of the phantom. The rigid registrations therefore yielded visible (though small) misalignment of the neck arteries, precluding assignment of 4D-flow velocity fields on the three corresponding outlets (\textit{BCA}, \textit{LCA}, \textit{LSA}). Experimental \textit{b1}, \textit{b2}, and \textit{b3} flow rates integrated from 2D cine PC-MRI velocities were therefore respectively prescribed at \textit{BCA}, \textit{LCA}, and \textit{LSA} with idealized parabolic velocity profiles. Lastly, experimental \textit{outlet} transducer pressures were used to generate the functional form of $P(t)$ prescribed at \textit{outlet}.

\subsubsection{Discrete FSI Formulation}
The semi-discrete formulation is constructed with the residual-based variational multiscale formulation. Considering a conforming mesh across $\Gamma_I$ immediately guarantees strong satisfaction of the kinematic coupling condition and weak satisfaction of the dynamic coupling condition.

Under a thin-walled assumption, the three-dimensional elastodynamic problem in $\Omega^s$ collapses to a two-dimensional problem posed on the fluid-solid interface $\Gamma_I$. Correspondingly, the outer wall $\Gamma_{W}$ collapses onto $\Gamma_{I}$, and annular surface $\Gamma^{i}_{\mathrm{cap}}$ collapses into a one-dimensional ring $\Gamma^{i}_{\mathrm{ring}} := \Gamma_{I} \cap \Gamma^{i}_{\mathrm{cap}}$. Let $\bm u^{w}_h$ be the membrane displacement on $\Gamma_I$ and $\mathcal S^{w}_{\bm u}$ be its trial solution space. Let $\mathcal S_{\bm v}^f$ and $\mathcal S_{p}^f$ denote the trial solution spaces for the fluid velocity and pressure; let $\mathcal V_{\bm v}^f$ and $\mathcal V_{p}^f$ be their corresponding test function spaces. The semi-discrete FSI formulation posed only in $\Omega^f$ on a stationary mesh is then stated as follows. Find $\bm y_h(t) := \left\lbrace \bm u^{w}_h(t), \bm v_h^f(t), p_h^f(t) \right\rbrace \in \mathcal S^w_{\bm u} \times \mathcal S_{\bm v}^f \times \mathcal S_{p}^f$ such that $\forall \left\lbrace \bm w_h^f, q_h^f\right\rbrace \in \mathcal V_{\bm v}^f \times \mathcal V_{p}^f$,
\begin{align*}
& \mathbf B_{\mathrm{k}} \left( \dot{\bm y}_h, \bm y_h \right) = \bm 0, &&  \\
& \mathbf B_{\mathrm{m}} \left( \bm w_h^f ;  \dot{\bm y}_h, \bm y_h \right) := \mathbf B^w_{\mathrm{m}} \left( \bm w_h^f ;  \dot{\bm y}_h, \bm y_h \right) + \mathbf B^f_{\mathrm{m}} \left( \bm w^f_h ;  \dot{\bm y}^f_h, \bm y^f_h \right)  = 0, \\
& \mathbf B_{\mathrm{c}}\left( q^f_h; \dot{\bm y}_h, \bm y_h \right) := \mathbf B_{\mathrm{c}}\left( q_h^f; \dot{\bm y}_h^f, \bm y_h^f \right) = 0,
\end{align*}
where 
\begin{align*}
& \mathbf B_{\mathrm{k}} \left( \dot{\bm y}_h, \bm y_h \right) := \frac{d\bm u^w_h}{dt} - \bm v^f_h, \qquad \mbox{ on } \Gamma_I, \\
& \mathbf B^w_{\mathrm{m}} \left( \bm w_h^f ;  \dot{\bm y}_h, \bm y_h \right) := \int_{\Gamma_I} \bm w_h^f \cdot \rho^s h^s \left( \frac{d \bm v_h^f}{d t} - \bm b^s \right) d\Gamma + \int_{\Gamma_I} h^s \bm \epsilon(\bm w_h^f) : \Big( \bm \sigma^s(\bm u_h^w) + \bm \sigma_0 \Big) d\Gamma \\ 
& \hspace{28mm} + \int_{\Gamma_I} c^s \bm w_h^f \cdot \bm v^{f} d\Gamma, \\
& \mathbf B^{f}_{\mathrm{m}} \left( \bm w_h^f ;  \dot{\bm y}_h^f, \bm y_h^f \right) := \int_{\Omega^f} \bm w_h^f \cdot \rho^f \left( \frac{\partial \bm v_h^f}{\partial t} + \bm v_h^f \cdot \nabla \bm v_h^f - \bm b^f \right) d\Omega - \int_{\Omega^f} \nabla \cdot \bm w_h^f p_h^f d\Omega \\
& \hspace{28mm} + \int_{\Omega^f} 2\mu^f  \bm \varepsilon(\bm w_h^f) : \bm \varepsilon(\bm v_h^f) d\Omega - \int_{\Omega^{f \prime}} \nabla \bm w_h^f : \left( \rho^f \bm v^{\prime} \otimes \bm v_h^f \right) d\Omega \\
& \hspace{28mm}  + \int_{\Omega^{f \prime}} \nabla \bm v_h^f : \left( \rho^f \bm w_h^f \otimes \bm v^{\prime} \right) d\Omega - \int_{\Omega^{f \prime}} \nabla \bm w_h^f : \left( \rho^f \bm v^{\prime} \otimes \bm v^{\prime} \right) d\Omega \\
& \hspace{28mm} - \int_{\Omega^{f \prime}} \nabla \cdot \bm w_h^f p^{\prime} d\Omega + \int_{\Gamma_h} \bm w_h^f \cdot \bm n^f P(t) d\Gamma - \int_{\Gamma_h}  \rho^f \beta \left(\bm v_h^f \cdot \bm n^f \right)_{-} \bm w_h^f \cdot \bm v_h^f d\Gamma, \\
& \mathbf B_{\mathrm{c}}\left( q_h^f; \dot{\bm y}_h^f, \bm y_h^f \right) := \int_{\Omega^f} q_h^f \nabla \cdot \bm v_h^f d\Omega  - \int_{\Omega^{f \prime}} \nabla q_h^f \cdot  \bm v^{\prime} d\Omega,
\end{align*}
and
\begin{align*}
& \bm v^{\prime} := -\bm \tau_{M} \left( \rho^f \frac{\partial \bm v_h^f}{\partial t} + \rho^f \bm v_h^f \cdot \nabla \bm v_h^f  + \nabla p_h^f - \mu^f \Delta \bm v_h^f - \rho^f \bm b^f \right), \quad p^{\prime} := -\tau_C \nabla \cdot \bm v_h^f, \\
& \bm \tau_{M} := \tau_M \bm I_3, \quad \tau_M := \frac{1}{\rho^f}\left( \frac{\mathrm C_{\mathrm T}}{\Delta t^2} + \bm v_h^f \cdot \bm G \bm v_h^f + \mathrm C_{\mathrm I} \left( \frac{\mu^f}{\rho^f} \right)^2 \bm G : \bm G \right)^{-\frac12}, \quad \tau_C := \frac{1}{\tau_M \textup{tr}\bm G}, \\
& G_{ij} := \sum_{k=1}^{3} \frac{\partial y_k}{\partial x_i} M_{kl} \frac{\partial y_l}{\partial x_j}, \quad \bm M = [ M_{kl} ] = \frac{\sqrt[3]{2}}{2}\begin{bmatrix}
2 & 1 & 1 \\
1 & 2 & 1 \\
1 & 1 & 2
\end{bmatrix}, \\
& \bm G : \bm G := \sum_{i,j=1}^{3} G_{ij} G_{ij}, \quad \textup{tr}\bm G := \sum_{i=1}^{3} G_{ii}, \quad \left( \bm v_h^f \cdot \bm n^f \right)_{-} := \frac{\bm v_h^f \cdot \bm n^f - |\bm v_h^f \cdot \bm n^f|}{2}.
\end{align*}

The prestress $\bm \sigma_0$ in $\mathbf B^w_{\mathrm{m}}$ that balances the in vivo blood pressure and viscous traction at the imaged configuration can be determined through a fixed-point algorithm \cite{Hsu2011, Lan2021}, and $\beta$ in the backflow stabilization term in $\mathbf B^{f}_{\mathrm{m}}$ is set to $0.2$. In the definition of $G_{ij}$, $\bm y = \left\lbrace y_i \right\rbrace_{i=1}^{3}$ are natural coordinates in the parent domain, and $\bm M$ is introduced to yield node-numbering-invariant definitions of $\tau_M$ and $\tau_C$ for simplex elements. $C_I$ and $C_T$ are taken to be $36$ and $4$ for linear tetrahedral elements.

The generalized-$\alpha$ method \cite{Lan2021, Liu2020a} is then applied for temporal discretization of the above semi-discrete FSI formulation. In this work, the time step size was uniformly selected as $T_p / 1000$.

\subsubsection{Numerical Simulation Strategies}
The fully discrete scheme was solved iteratively via a predictor multi-corrector algorithm. To reduce the size of the associated linear system and enable effective block preconditioning \cite{Lan2021, Liu2020}, we leveraged the special block structure of the fully consistent tangent matrix to develop a segregated algorithm. Our block preconditioner was previously shown to exhibit enhanced robustness and scalability as compared to alternative preconditioners in applications spanning hyperelasticity, viscous fluids, and FSI \cite{Lan2021, Liu2019}.

To reflect the steady configuration captured by the 3D SPGR scan, we initialized our simulation in the following steps. A rigid-walled CFD simulation was first performed with a prescribed steady inflow of $71.2$ mL/s to determine the corresponding fluid traction $\bm h^f$. Resistance boundary conditions were prescribed at all four outlets, tuned to achieve the experimental outflow distribution ($18.0\%$ \textit{BCA}, $4.9\%$ \textit{LCA}, $3.8\%$ \textit{LSA}, $73.3\%$ \textit{outlet}) and mean experimental pressures (\textit{inlet} $56$ mm Hg, \textit{outlet} $55$ mm Hg). The resulting $\bm h^f$ was then used in the fixed-point algorithm to determine $\bm \sigma_0$. With the vascular wall appropriately prestressed, we performed a steady-state FSI simulation prescribed with steady diastolic boundary conditions corresponding to $t=0$, from which we initialized our pulsatile simulation. Three cardiac cycles were simulated to ensure convergence to a limit cycle, and only the third cycle was analyzed.

Comparisons between experimental and simulated velocity fields were made possible by sampling the 4D-flow velocity fields onto the computational mesh. With the same TTF approach discussed in Section \ref{subsec:experimental-analysis}, we computed the numerical PWV using simulated pressure waveforms at the same $50$ normal slices. We note that in contrast to the use of flow waveforms to compute the experimental PWV, pressure waveforms were instead used to compute the numerical PWV due to their uniformity in waveform shape.

\begin{figure}
\begin{center}
\begin{tabular}{c}
\includegraphics[angle=0, trim=88 160 70 155, clip=true, scale=0.74]{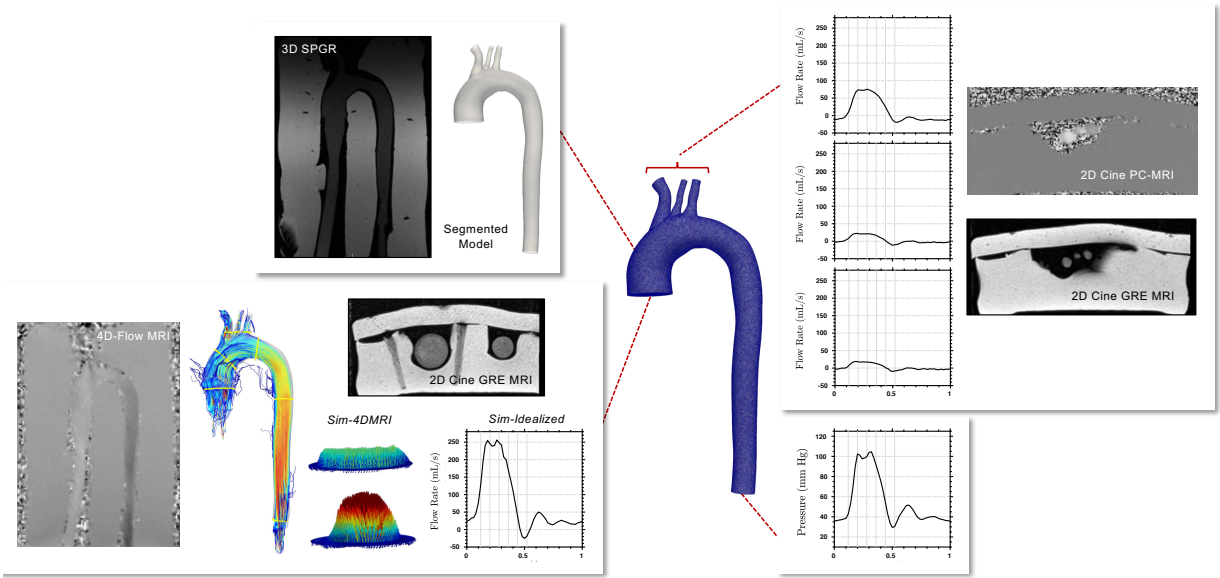}
\end{tabular}
\end{center}
\caption{Schematic of numerical simulation methods incorporating data from four MRI sequences. The anatomical model was segmented from the 3D SPGR scan. Velocities from 2D cine PC-MRI were integrated over lumen areas from 2D cine GRE MRI to generate the volumetric flow rates prescribed at \textit{BCA}, \textit{LCA}, and \textit{LSA} with idealized parabolic velocity profiles. The \textit{outlet} pressure measured by a pressure transducer was prescribed. Velocities from 4D-flow MRI were masked by lumen contours from 2D cine GRE MRI to generate the velocity profiles prescribed at \textit{inlet} in \textit{Sim-4DMRI}. These 4D-flow velocities were integrated to generate the volumetric flow rates prescribed at \textit{inlet} with parabolic velocity profiles in \textit{Sim-Idealized}.}
\label{fig:simulation-methods}
\end{figure}

\subsubsection{Alternative Numerical Boundary Conditions}
In addition to the simulation outlined above, which we refer to as \textit{Sim-4DMRI} (\textbf{Figure \ref{fig:simulation-methods}}), we also alternatively prescribed an idealized parabolic inlet velocity profile in a separate simulation termed \textit{Sim-Idealized}. Experimental flow rates integrated from the masked 4D-flow velocities were thus prescribed at \textit{inlet} with parabolic velocity profiles, and all other aspects of \textit{Sim-Idealized} were kept consistent with \textit{Sim-4DMRI}. 

We further performed three supplemental simulations to compare the effects of the in-plane and clamped boundary conditions on $\Gamma^{i}_{\mathrm{ring}}$ and assess the effect of reducing the viscous damping constant $c^s$ applied over $\Gamma_I$. Noting that \textit{Sim-4DMRI} was performed with in-plane motion and  $c^s = 3.0 \times 10^5$ g/(cm$^2 \cdot$ s), the supplemental simulations enabled investigation of the following four combinations: (i) in-plane motion, $c^s = 3.0 \times 10^5$ (\textit{Sim-4DMRI}); (ii) in-plane motion, $c^s = 3.0 \times 10^3$; (iii) clamped, $c^s = 3.0 \times 10^5$; and (iv) clamped, $c^s = 3.0 \times 10^3$. Experimental 4D-flow velocity profiles were prescribed at \textit{inlet} across these four simulations. Removing the viscous damping altogether would have allowed us to decouple the effects of these boundary conditions on $\Gamma^{i}_{\mathrm{ring}}$ and $\Gamma_I$, but we have found that for practical meshes of reasonable cost, enabling in-plane motion without prescribing any external tissue support results in arbitrary translation of $\Gamma^{i}_{\mathrm{ring}}$ nodes due to inherent mesh asymmetry.

\begin{figure}
\begin{center}
\begin{tabular}{c}
\includegraphics[angle=0, trim=210 90 30 90, clip=true, scale=1.2]{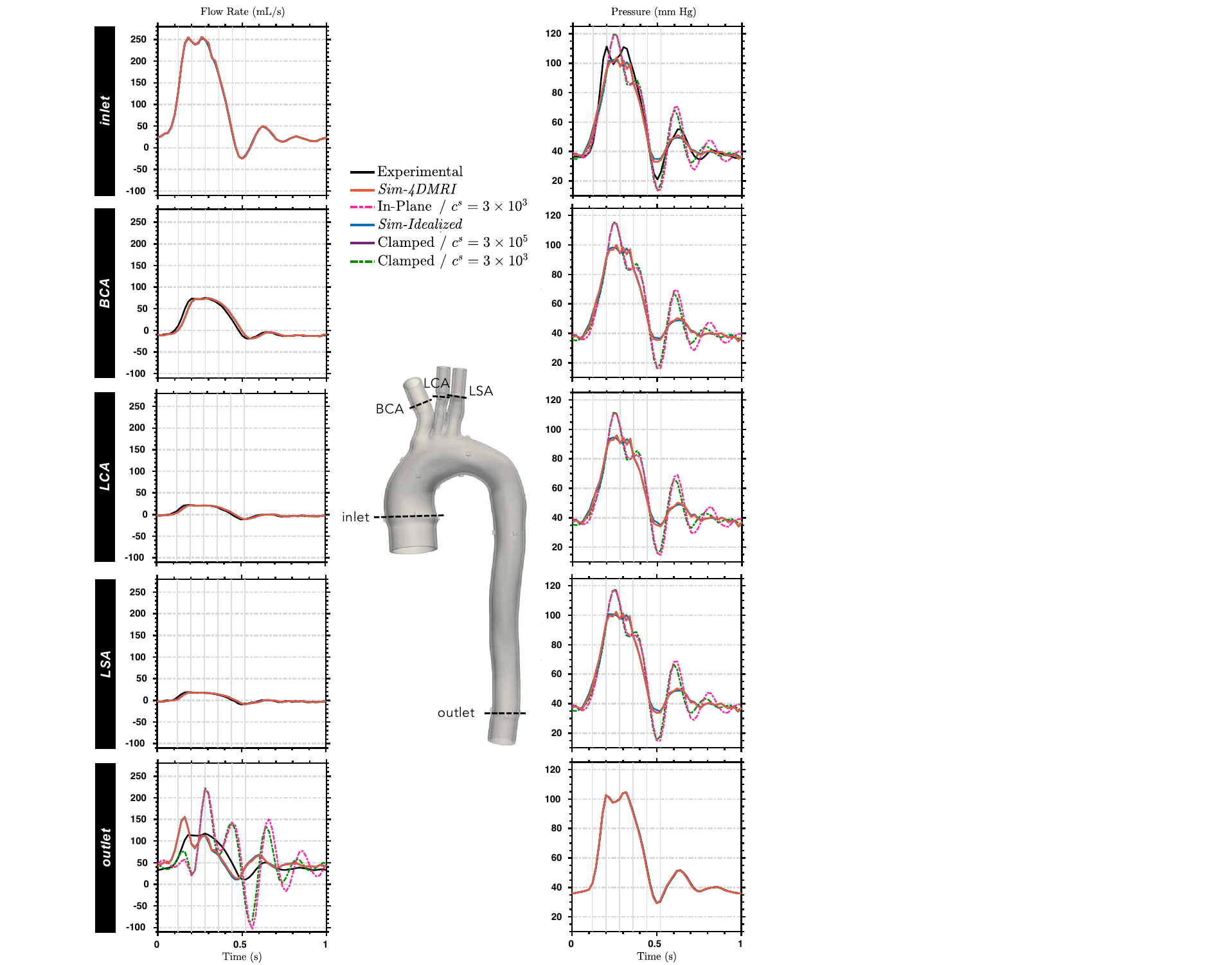}
\end{tabular}
\end{center}
\caption{Comparison of experimental and simulated volumetric flow rates and pressures over time at the five caps. In both \textit{Sim-4DMRI} and \textit{Sim-Idealized}, in-plane motion of the wall boundary rings is enabled, and the prescribed damping constant is $c^s = 3 \times 10^5$ g/(cm$^2 \cdot$ s).} 
\label{fig:flow-pres}
\end{figure}

\begin{figure}
	\begin{center}
	\begin{tabular}{c}
\includegraphics[angle=0, trim=245 173 30 175, clip=true, scale=1.4]{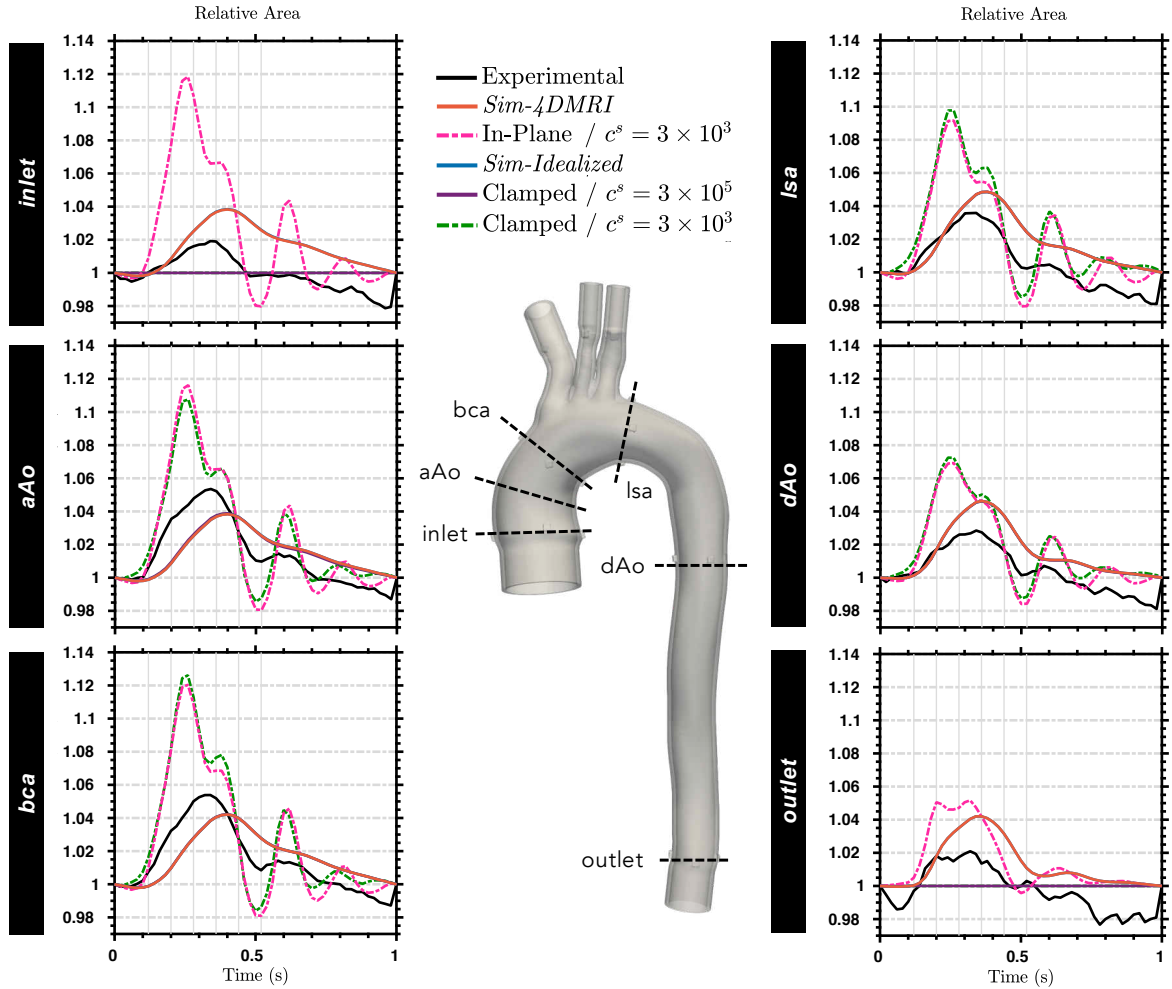}
\end{tabular}
\end{center}
\caption{Comparison of experimental and simulated relative areas over time at the six 2D scan landmarks along the aortic arch. In both \textit{Sim-4DMRI} and \textit{Sim-Idealized}, in-plane motion of the wall boundary rings is enabled, and the prescribed damping constant is $c^s = 3 \times 10^5$ g/(cm$^2 \cdot$ s). We note that the experimentally measured areas do not exhibit periodicity, displaying a sharp jump from $t=0.98$ s to the artificially repeated $t=0.0$ value for $t=1.0$ s.} 
\label{fig:rel-area}
\end{figure}

\section{Results}
Comparisons between simulated and experimentally measured volumetric flow rates, pressures, and relative areas over the cardiac cycle are shown in \textbf{Figures \ref{fig:flow-pres} and \ref{fig:rel-area}}. Whereas the experimental flow rates plotted for \textit{inlet} and \textit{outlet} were computed from 4D-flow velocities, those plotted for \textit{BCA}, \textit{LCA}, and \textit{LSA} were computed from 2D cine PC-MRI velocities, given the 4D-flow MRI registration discrepancies discussed in Section \ref{subsec:bcs}. We focus first on \textit{Sim-4DMRI} and \textit{Sim-Idealized}. As expected, flow rate discrepancies from the experimental results were observed only for \textit{outlet}, the only cap for which a Neumann condition was prescribed. While exact mass conservation would not be expected either experimentally or numerically, the experimental flow rates exhibited a mean loss of $5.01$ mL/s from the inflow ($77.8$ mL/s); by comparison, \textit{Sim-4DMRI} and \textit{Sim-Idealized} exhibited mean gains of only $0.0671$ mL/s and $0.0765$ mL/s, respectively, from the inflow ($77.8$ mL/s).

Relative to the experimental \textit{inlet} pressure, the corresponding \textit{Sim-4DMRI} and \textit{Sim-Idealized} pressures were lower in systole and higher in diastole, suggesting larger numerical compliance. The experimental \textit{outlet} pressure was numerically prescribed, and no experimental pressure measurements were available for \textit{BCA}, \textit{LCA}, or \textit{LSA}. This increased numerical compliance, which was further supported by the larger simulated relative area changes in $4$ of the $6$ landmarks along the aortic arch, can be attributed to the adopted linear elasticity with a tangential Young's modulus taken at the mean pressure. Closer agreement would be expected with future extensions to nonlinear models.

Turning our attention to the three supplemental simulations, we observed that the simulation with the same damping constant $c^s = 3 \times 10^5$ g/(cm$^2 \cdot$ s) as \textit{Sim-4DMRI} but with clamped wall boundary rings yielded flow rates, pressures, and relative areas that were practically indistinguishable from those of \textit{Sim-4DMRI}, with the exception of the constant \textit{inlet} and \textit{outlet} areas. On the other hand, the two simulations with the reduced $c^s = 3 \times 10^3$ g/(cm$^2 \cdot$ s) exhibited significant oscillations in all non-prescribed flow rates and pressures and all areas at non-clamped landmarks.

\begin{figure}
\begin{center}
\begin{tabular}{c}
\includegraphics[angle=0, trim=70 95 30 95, clip=true, scale=0.70]{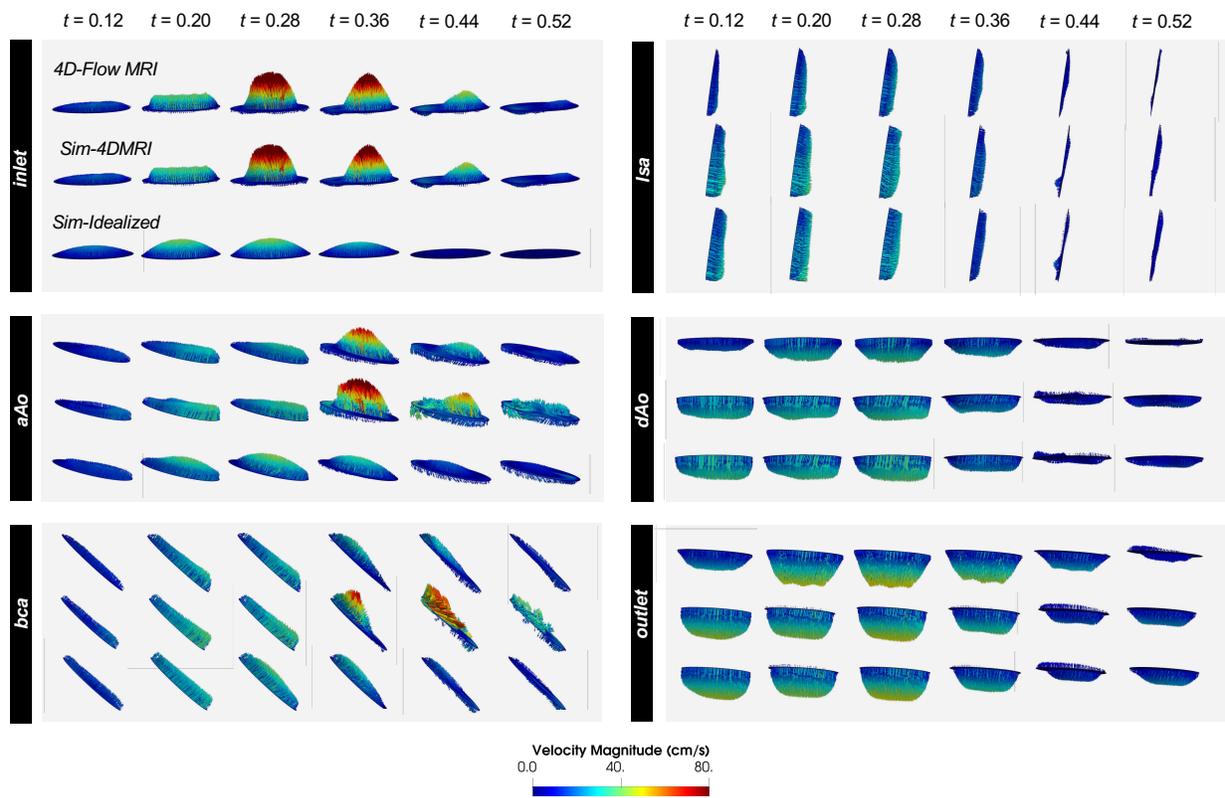}
\end{tabular}
\end{center}
\caption{4D-flow MRI (top), \textit{Sim-4DMRI} (middle), and (C) \textit{Sim-Idealized} (bottom) velocity profiles at $6$ evenly spaced temporal frames spanning the systolic phase of the cardiac cycle. All six 2D scan landmarks along the aortic arch are included.} 
\label{fig:slice-velo-glyphs}
\end{figure}

\begin{figure}
\begin{center}
\begin{tabular}{c}
\includegraphics[angle=0, trim=62 90 30 90, clip=true, scale=1.0]{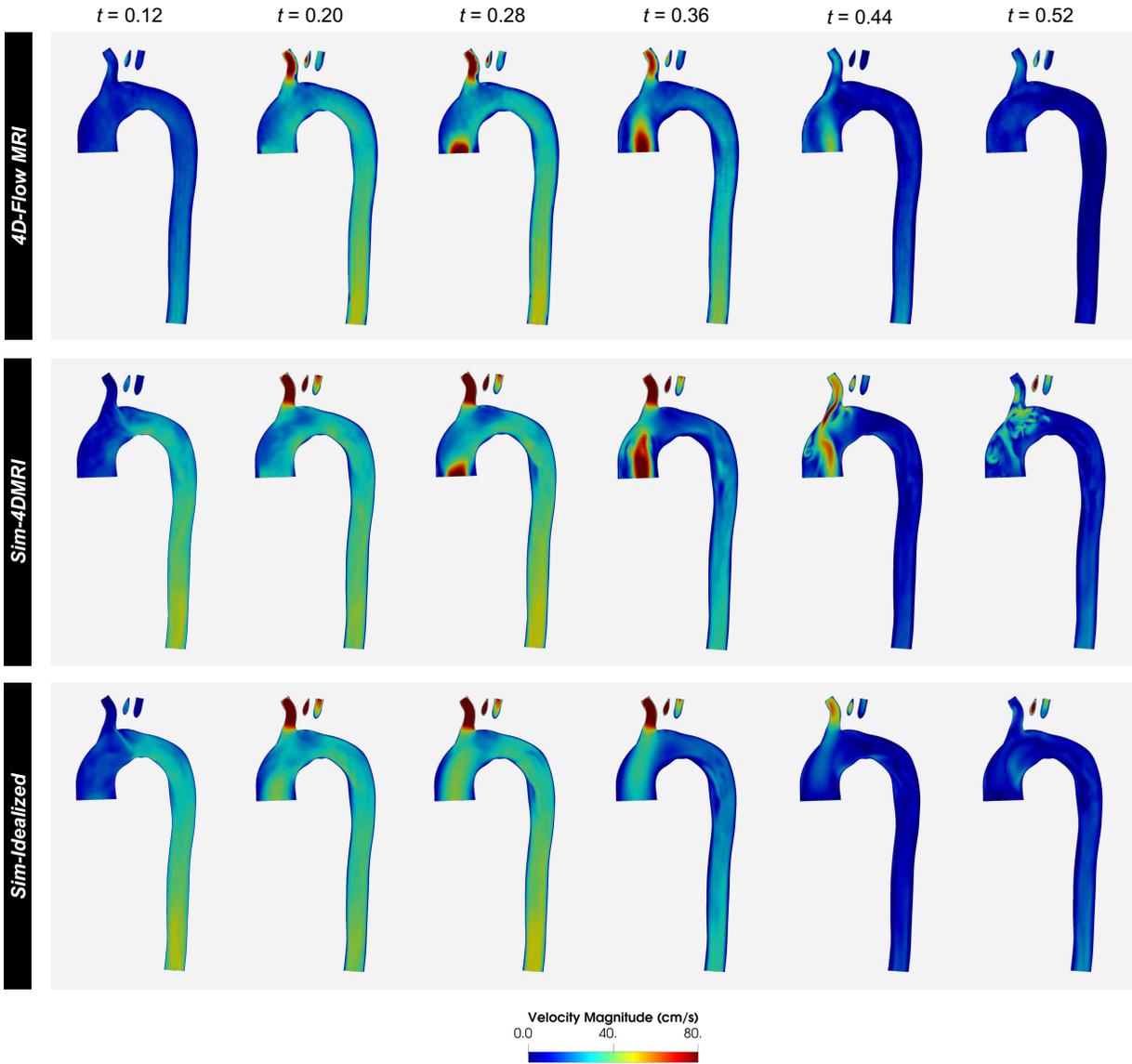}
\end{tabular}
\end{center}
\caption{4D-flow MRI (top), \textit{Sim-4DMRI} (middle), and \textit{Sim-Idealized} (bottom) velocity profiles on a sagittal plane at $6$ evenly spaced temporal frames spanning the systolic phase of the cardiac cycle.} 
\label{fig:sagittal-velo}
\end{figure}

\begin{figure}
\begin{center}
\begin{tabular}{c}
\includegraphics[angle=0, trim=145 90 185 90, clip=true, scale=1.3]{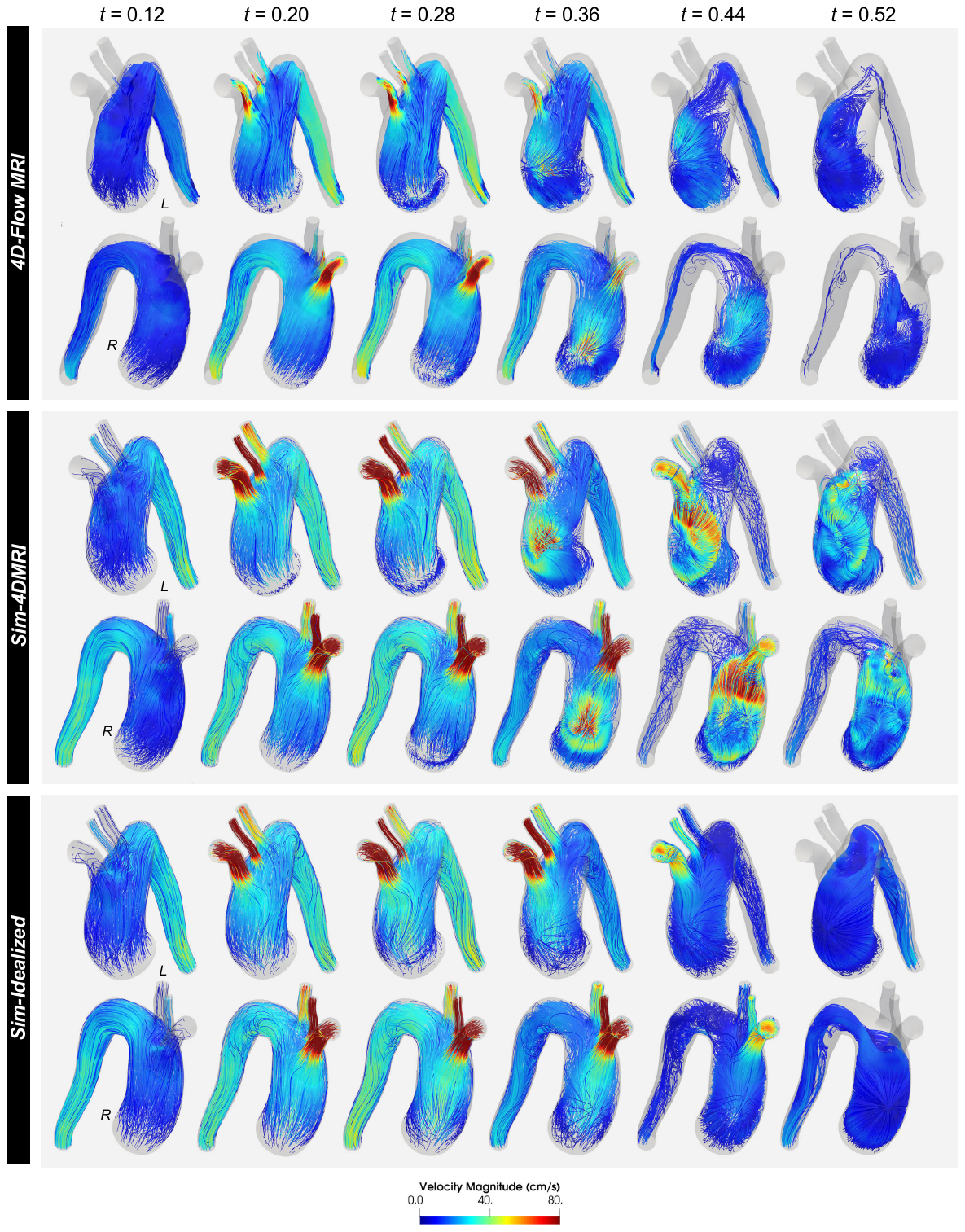}
\end{tabular}
\end{center}
\caption{Streamlines computed from 4D-flow MRI (top), \textit{Sim-4DMRI} (middle), and \textit{Sim-Idealized} (bottom) velocities at $6$ evenly spaced temporal frames spanning the systolic phase of the cardiac cycle. The left (L, top) and right (R, bottom) lateral sides are annotated.} 
\label{fig:streamlines}
\end{figure}

We investigated the experimental 4D-flow MRI and simulated velocity profiles (\textbf{Figure \ref{fig:slice-velo-glyphs}}) at $6$ evenly spaced temporal frames spanning systole (annotated in \textbf{Figures \ref{fig:flow-pres} and \ref{fig:rel-area}}), beginning with acceleration at $t=0.12$ s, peak systole from $t=0.20$ to $t=0.28$ s, and ending with deceleration from $t=0.36$ to $t=0.52$ s. All six 2D scan landmarks along the aortic arch were included. Focusing only on \textit{Sim-4DMRI}, we observed close agreement in the profile shapes in all frames leading up to peak systole albeit the higher simulated velocities in the descending aorta, as expected given the higher simulated flow rates seen in \textbf{Figure \ref{fig:flow-pres}}. Ascending aortic velocity magnitudes in the deceleration phase, however, revealed that the simulation captured stronger forward flow as well as stronger reverse flow along the exterior curve of the ascending aorta. These differences were similarly observed from sagittal velocity fields at the same temporal frames (\textbf{Figure \ref{fig:sagittal-velo}}), noting again that no comparisons could be drawn for the misaligned neck arteries. Barring the higher velocity magnitudes in the descending aorta at $t=0.12$ s, our simulation almost exactly reproduced the experimental velocity fields leading up to peak systole. We also observed close agreement during fluid deceleration with regard to the shape and angle of the jet, including its curved tip near the brachiocephalic artery take-off.

To comprehend the 3D flow behavior, we turned our attention to streamlines (\textbf{Figure \ref{fig:streamlines}}). In both \textit{Sim-4DMRI} and the experiment, flow near the inlet reversed along the proximal aortic wall at the onset of fluid deceleration following peak systole at $t=0.28$ s. This reversed flow continued to grow in strength until formation of a vortex at $t=0.40$ s along the left lateral wall. Despite the significantly lower velocity magnitudes in the 4D-flow data, we observed notable agreement in the vortex shape and location in the final two highlighted frames. We do, however, note the presence of three additional smaller vortices in \textit{Sim-4DMRI} that were absent from the experiment: one along the right lateral wall (formed at $t=0.46$s), and two on either side of the sagittal plane (formed at $t=0.50$ s). Whereas the single vortex in the experimental data persisted until the end of the cardiac cycle, the presence of multiple interacting vortices in \textit{Sim-4DMRI} caused the four initial vortex structures to progressively break up into increasing numbers of smaller vortices.

In contrast to these \textit{Sim-4DMRI} results, velocity fields from \textit{Sim-Idealized} failed to capture key salient features of the experimental fields, including the shape and angle of the jet. Furthermore, the ascending aortic velocity magnitudes were much too low throughout systole, the onset of flow reversal was delayed until $t=0.40$ s, and vortex formation was similarly delayed until the end of systole at $t=0.56$ s.

\begin{figure}
	\begin{center}
	\begin{tabular}{c}
\includegraphics[angle=0, trim=128 230 155 120, clip=true, scale=0.9]{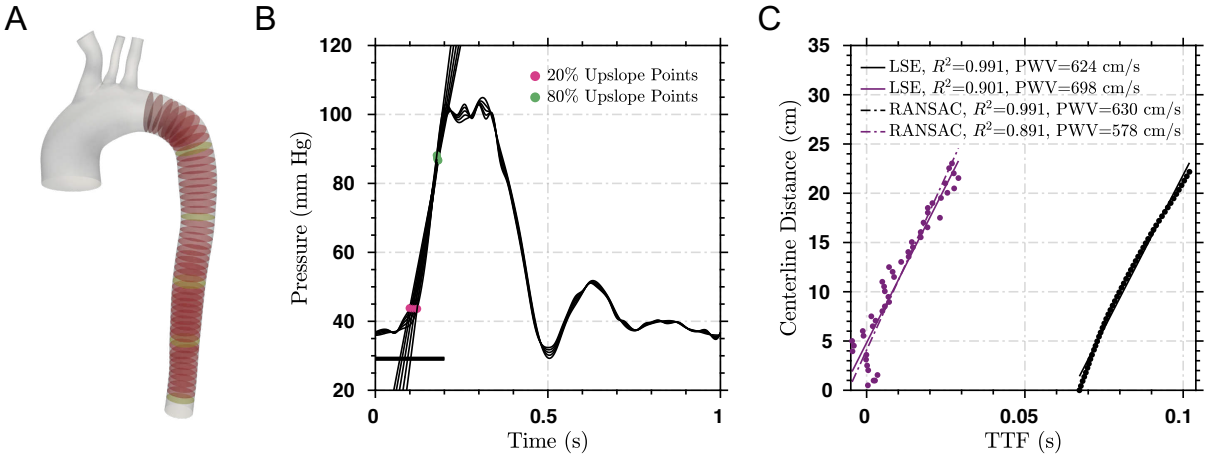}
\end{tabular}
\end{center}
\caption{Numerical pulse wave velocity (PWV) computation for \textit{Sim-4DMRI} using (A) $50$ equidistant normal slices along the descending aorta. (B) Simulated pressure waveforms and lines used to determine times-to-foot (TTFs) are plotted for the five representative yellow slices. (C) Experimental (purple) and numerical (black) PWV estimates determined as slopes of linear regressions, performed with either least squares error (LSE) or Random Sample Consensus (RANSAC), relating TTF to the distance along the descending aortic centerline. The time delay seen in the numerical results relative to the experimental results is a consequence of the use of pressure rather than flow waveforms.} 
\label{fig:dao-pwv}
\end{figure}

In computing the experimental PWV, the presence of numerous outlier TTF points yielded $698$ cm/s with an LSE regression but $578$ cm/s with a RANSAC regression. By comparison, LSE and RANSAC regressions for \textit{Sim-4DMRI} yielded similar PWV estimates of $624$ cm/s and $630$ cm/s, respectively, both with equally high coefficients of determination $R^2$ indicating excellent fits (\textbf{Figure \ref{fig:dao-pwv}}). Importantly, our simulated PWV values fell within the experimental bounds.

\section{Discussion}
Our study represents the first to validate a cardiovascular FSI formulation against an in vitro flow circuit involving a compliant patient-specific vascular phantom. Novel aspects spanning both the experimental and numerical fronts include (i) rapid, cost-effective, and repeatable PolyJet photopolymerization of a compliant aortic model; (ii) mechanical characterization of the phantom for FSI parameter estimation; (iii) the use of multiple MRI sequences to achieve high-resolution anatomical and hemodynamic data; (iv) in-plane motion of inlet and outlet surfaces to circumvent nonphysiological clamping; (v) external tissue support to model the embedding gel block; and (vi) vascular wall prestressing to reflect imaging under experimental steady flow conditions.

When Moireau et al. \cite{Moireau2012} first proposed external tissue support, they also found it effective at eliminating rapid wall oscillations observed with CMM \cite{Figueroa2006}, which have long been hypothesized to arise from clamping-induced wave reflections. Nonetheless, \textbf{Figures \ref{fig:flow-pres} and \ref{fig:rel-area}} clearly indicate that even with in-plane motion of $\Gamma^i_{\mathrm{ring}}$ nodes, viscous damping beyond some threshold level was necessary to suppress oscillations in our study. To better understand the source of these oscillations, we additionally replaced the flow rate and pressure boundary conditions at the outlets with Windkessel models; no oscillations were observed even when clamping  $\Gamma^i_{\mathrm{ring}}$ nodes in the absence of external tissue support. Prescribing the resulting \textit{inlet}, \textit{BCA}, \textit{LCA}, and \textit{LSA} flow rates and \textit{outlet} pressure from this Windkessel simulation in a subsequent simulation also produced no oscillations. Together, these observations suggested that the oscillations requiring damping in our study were driven by the assignment of experimental waveforms.

Despite discrepancies in the velocity fields during deceleration, our simulation captured key salient features of the 4D-flow MRI data, including the (i) region and trajectory of reverse flow following peak systole, (ii) shape and angle of the jet, and (iii) shape and location of the primary vortex. On the other hand, \textit{Sim-Idealized} exhibited poor agreement with the 4D-flow MRI data in numerous aspects including the ascending aortic velocity magnitudes, shape and angle of the jet, duration of flow reversal, and time of vortex formation. These discrepancies highlight the importance of assigning non-idealized inlet velocity profiles. 

Previous in vitro and in vivo validation studies \cite{Kung2011a, Biglino2015, Saitta2019} have similarly reported larger velocity field discrepancies during deceleration, with simulations exhibiting complex secondary flow features that were absent from the 4D-flow MRI data. These discrepancies in complex flow structures can partially be explained by the many issues inherent to 4D-flow MRI. In particular, despite the many corrections applied to the acquired velocity fields either during or following image reconstruction to reduce phase offsets, the resulting 4D-flow MRI velocity fields are not completely divergence-free, signaling a lack of conservation of mass. Solenoidal filtering and related divergence-free methods \cite{Schiavazzi2014} could be explored in the future. Furthermore, image acquisition in k-space over extended durations of time yields images that are temporally averaged over tens of cardiac cycles, and furthermore averaged within each temporal bin corresponding to the temporal resolution ($0.02$ s here). Boundary voxels capturing both the fluid and surrounding wall are also subject to the partial volume effect \cite{GonzalezBallester2000}. The combined effect of partial voluming and temporal averaging over a deforming domain presents challenges for deriving accurate boundary conditions reflective of the in vivo or in vitro conditions to be prescribed in simulations. Additional sources of error existed in our study, as indicated by the lack of conservation of mass, absence of periodicity in luminal areas, and the large number of outliers yielding regression-dependent PWV estimates. Moreover, the misalignment of the neck arteries hindered our ability to assign non-idealized velocity profiles at the corresponding \textit{BCA}, \textit{LCA}, and \textit{LSA} outlets. The resulting parabolic profiles and limited unidirectional velocity encoding of 2D cine PC-MRI likely introduced discrepancies that were propagated into the flow domain.

Overall, close agreement in pressures, lumen area changes, pulse wave velocity, early systolic velocities, and late systolic flow structures validates our numerical methodology, including the RUC formulation, in-plane deformation at model inlets and outlets, viscoelastic external tissue support, and vascular tissue prestressing. Together, our suite of computationally efficient FSI techniques offers a platform for investigating vascular disease initiation, progression, and treatment.

\section*{Acknowledgments}
This work was supported by the National Institutes of Health [grant numbers 1R01HL121754, 1R01HL123689, R01EB01830204], National Natural Science Foundation of China [grant number 12172160], and Guangdong-Hong Kong-Macao Joint Laboratory for Data-Driven Fluid Mechanics and Engineering Applications [grant number 2020B1212030001]. Ingrid S. Lan was supported by the National Science Foundation (NSF) Graduate Research Fellowship and Stanford Graduate Fellowship in Science and Engineering. Computational resources were provided by the Stanford Research Computing Center and Extreme Science and Engineering Discovery Environment supported by NSF [grant number ACI-1053575].


\bibliographystyle{abbrv}
\bibliography{preprint_RUC_validation}

\end{document}